# Magnetic hyperthermia experiments with magnetic nanoparticles in clarified butter oil and paraffin: a thermodynamic analysis

*Adriele A. de Almeida[1], Emilio De Biasi[1,2], Marcelo Vasquez Mansilla[1], Daniela P. Valdés[1,2], Horacio E. Troiani[2,3], Guillermina Urretavizcaya[4], Teobaldo E. Torres[1], Luis M. Rodríguez[1], Daniel E. Fregenal[5], Guillermo C. Bernardi[1], Elin L. Winkler[1,2], Gerardo F. Goya[6,7], Roberto D. Zysler[1,2], Enio Lima Jr.[1,\*]*

[1] Instituto de Nanociencia y Nanotecnología, CNEA-CONICET, Centro Atómico Bariloche, S. C. de Bariloche, 8400, RN, Argentina.

[2] Instituto Balseiro, CNEA-UNCuyo, Centro Atómico Bariloche, S. C. de Bariloche, 8400, RN, Argentina.

[3] Lab. de Caracterización de Materiales y Óxidos No-Estequiométricos, Gerencia de Investigación Aplicada, Centro Atómico Bariloche, S. C. de Bariloche, 8400, RN, Argentina.

[4] Consejo Nacional de Investigaciones Científicas y Técnicas (CONICET) e Instituto Balseiro, Centro Atómico Bariloche, S. C. de Bariloche, 8400, RN, Argentina.

[5] Centro Atómico Bariloche, Bustillo 9500, San Carlos de Bariloche, 8400, Río Negro, Argentina

[6] Dep. de Física de la Materia Condensada, Universidad de Zaragoza, Mariano Esquillor s/n, Zaragoza, 50018, Spain.

[7] Instituto de Nanociencia de Aragón, Universidad de Zaragoza, Mariano Esquillor s/n, Zaragoza, 50018, Spain.

* author to whom correspondence: lima@cab.cnea.gov.ar

ABSTRACT. In Specific Power Absorption (SPA) models for Magnetic Fluid Hyperthermia (MFH) experiments, the magnetic relaxation time of the nanoparticles (NPs) is known to be a fundamental descriptor of the heating mechanisms. The relaxation time is mainly determined by the interplay between the magnetic properties of the NPs and the rheological properties of NP's environment. Although the role of magnetism in MFH has been extensively studied, the thermal properties of the NPs medium and their changes during of MFH experiments have been so far underrated. Here, we show that $Zn_xFe_{3-x}O_4$ NPs dispersed through different with phase transition in the temperature range of the experiment: clarified butter oil (CBO) and paraffin. These systems show non-linear behavior of the heating rate within the temperature range of the MFH experiments. For CBO, a fast increase at ~ 306 K associated to changes in the viscosity ($\eta(T)$) and specific heat ($c_P(T)$) of the medium below and above its melting temperature. This increment in the heating rate takes place around 318 K for paraffin. Magnetic and morphological characterizations of NPs together with the observed agglomeration of the nanoparticles above 306 K indicate that the fast increase in MFH curves could not be associated to a change in the magnetic relaxation mechanism, with Néel relaxation being dominant. In fact, successive experiment runs performed up to temperatures below and above the CBO melting point resulted in different MFH curves due to agglomeration of NPs driven by magnetic field inhomogeneity during the experiments. Similar effects were observed for paraffin. Our results highlight the relevance of the NPs medium's thermodynamic properties for an accurate measurement of the heating efficiency for in vitro and in vivo environments, where the thermal properties are largely variable within the temperature window of MFH experiments.

INTRODUCTION

In a magnetic fluid hyperthermia (MFH) experiment, the relaxation time of the magnetic moment ($\tau$) of nanoparticles (NPs) is a key factor for heating mechanisms operating behind Specific Power Absorption (SPA) determination. In a single domain NP, the effective results from the interplay of two relaxation mechanisms: the Brownian rotation against the viscous forces expressed by $\tau_B = 3\eta V_{hyd}/k_B T$, where $\eta$ is the viscosity of the medium, $V_{hyd}$ is the hydrodynamic volume of the NP and $k_B T$ is the thermal energy, and the Néel relaxation of the magnetic moment with a time given by $\tau_N = \tau_0 \exp(K_{eff} V/k_B T)$, where $\tau_0$ is the characteristic attempt time and $K_{eff} V$ is the effective anisotropy energy barrier of the NP given by its effective anisotropy constant $K_{eff}$ and its volume V [1]. The dominant mechanism will be the one with shorter relaxation time. For MFH applications, the most adequate systems are those in which $\tau_N$ dominates [2-5] because the heating power depends only on the intrinsic magnetic properties of the NPs and it is independent of the (variable) viscosity of different biological media. Therefore, the precise heat dose delivered is independent of the specific tissue or organ targeted [6-10]. There is a growing consensus on the fact that magnetic NPs for MFH should be fabricated in this way, and optimization done by adjusting its $K_{eff}V$.

Only few studies have been reported so far evaluating the variability of SPA expected when the NPs are dispersed in a medium with a temperature-dependent viscosity ($\eta(T)$) or specific heat ($c_p(T)$) within the temperature range measured [11]. If the medium presents a strong reduction of $\eta(T)$ in the temperature range of the experiment, for instance due to the melting, it results in a shorter $\tau_B$, which may lead to a change in the dominant mechanism acting behind the heating process. However, even when the magnetic relaxation ($\tau_N$) still dominating, changes in the properties of the medium may affect the SPA measured for the system. Changes in the viscosity may affect the spatial arrangement of the NPs, allowing the alignment of their magnetic easy axis with the external field gradient and yielding the formation of multi-particle structures with a very different SPA response [12-14]. In the case of a phase transition such as melting, the latent heat (L) also impacts on the SPA measured.

The paraffin and the clarified butter oil (CBO) are interesting media in order to study this subject. Both present a phase transition in the temperature range of the hyperthermia experiments. The paraffin is a simpler medium in terms of composition, usually a mixture of linear *n*-alkanes ($CH_3$-$(CH_2)_n$-$CH_3$), with a phase transition, depending on the number of carbon atoms, between 293 K – 295 K ($C_{16}$-$C_{18}$) and 339 K – 341 K ($C_{21}$-$C_{50}$) [15]. It has a latent heat of 121.9 J/g, a thermal conductivity of ~ 0.21 – 0.24 W/mK just below the melting point [15] and a dynamic viscosity around $\eta \sim 2.1$ mPa.s at the melting temperature of $T_M = 332$ K [16]. In another way, the CBO is a more complex medium obtained by extraction from heated butter oil, which could be interesting inter of potential uses of nanoparticles in applications of MFH. It is composed by 99 - 99.5 % of the fatty acids palmitic, oleic, myristic and stearic [17]. The use of CBO avoids the presence of water droplets that can affect the solidification process of the fat, and consequently the physicochemical properties of the crystal network [18]. Physical properties of CBO are well known, with $T_M$=305.2 K [19] (compared to $T_M$=308.0 K reported for the butter [20]). The viscosity of CBO decreases at the melting point by a factor of $10^3$, from some Pa.s to tens of mPa.s [21]. Also the thermal properties, like the $c_p$, vary with the temperature, being sensitive to changes in its physical state close to $T_M$ [22]. In fact, there are distinct regions of phase transition related to the melting: the Low-Melting Fraction (LMF), around 280 K, Medium-Melting Fraction (MMF), around 290 K, and a High-Melting Fraction (HMF), around 310 K, where the temperature range depends on the heating rate [23]. Its thermal conductivity slightly before the melting point is estimated in 0.22 W/m.K [21].

In this work, we study the SPA of Zinc substituted ferrite with general formula $Zn_xFe_{3-x}O_4$ in CBO and paraffin, two media with a phase transition with a strong reduction of the viscosity in the temperature range of MFH experiments. The first medium presents more complex composition than the second one, which may affect the response of the system during the hyperthermia experiments and should be taken into account if considering applications of MFH. We synthesized $Zn_xFe_{3-x}O_4$ nanoparticles with different mean diameter (<d>) and $K_{eff}$. The variations in $K_{eff}$ and <d> of the NPs result in different values of SPA, consequently to different measured heating rates. In addition, the incorporation of Zn in the ferrite structure reduces the anisotropy of the system allowing the heating at lower temperatures, where the viscosity of the medium is high. With this set of samples, we studied the SPA in the two media with variable viscous during the temperature increment in the MFH experiment for samples with different initial SPA values, different shapes and different sizes, which could be important to the effects of agglomeration observed at higher temperatures. A non-usual thermal variation of the heating rate is observed in MFH experiments for all samples, assigned to changes in the properties of the medium with temperature together with a spatial rearrangement of the NPs. Finally, irreversibility in the MFH response after successive experiments (runs) is observed for samples with higher NPs concentration as resulting from agglomeration above the melting of CBO and paraffin.

METHODS

$Zn_xFe_{3-x}O_4$ nanoparticles were produced by the thermal decomposition at high temperature of the Fe(III) and Zn(II) acetylacetonates in different solvents (Benzyl ether and trioctylamine) using the surfactants oleic acid and oleylamine. The method was adapted from the procedure described by Lohr *et al.* [24], trying to avoid the presence of Wüstite phase. The composition of each sample, with general formula $Zn_xFe_{3-x}O_4$ with x varying in the 0.1 - 0.2 range, was controlled by the precursors ratio used. The as-prepared NPs are hydrophobic, coated with oleic acid. In order to understand the dominant relaxation mechanisms in MFH experiments we performed a morphological characterization and correlated the observed magnetic properties with the phase composition of the NPs.

The composition of the as-prepared nanoparticles was determined by particle-induced X-ray emission (PIXE) [25] using a 3 MeV $H^+$ beam from a NEC 5SDH 1.7 MV tandem accelerator with a NEC RC43 end-station. For this, NPs were washed several times with acetone in order to reduce the oleic acid and oleylamine amounts remaining from the synthesis, and dried. The resulting powder was conditioned over a carbon tape for PIXE measurement and the results were analyzed with the software GUPIX [26]. As PIXE technique is very sensitive, signals from other elements, arising from the sample handling and conditioning, are observed: carbon, aluminum, sodium, calcium and silicon. Therefore, the composition of each sample was focused only in determine the ratio between the zinc and iron amounts ([Zn] and [Fe], respectively), resulting in x=0.11, 0.12, 0.19 and 0.21 (see Supporting Information S1). Samples were labeled according to this.

Size and diameter dispersion of as-prepared NPs were determined from transmission electron microscopy (TEM) obtained in a Philips CM200 microscope operating at 200 kV. TEM specimens were prepared by dropping a solution of NPs dispersed in toluene onto a copper grid coated with amorphous carbon layer. The diameter histograms were built up by measuring the diameter of at least 300 NPs, and the mean diameter <d> and dispersion σ were obtained by fitting it with a lognormal distribution. XRD profile of sample x = 0.21 as representative of the samples is shown in Supporting Information S2. As observed, diffraction peaks observed can be addressed to the cubic magnetite phase (JCPDS card 19-0629). No evidence of the ZnO phase is observed in the limit of our background and peak width, as indicated by the absence of diffraction in the positions expected for the most intense peaks of the hexagonal ZnO phase (JCPDS card 36-1451). In this, we assume that the Zn obtained in the PIXE analysis is incorporated in the cubic ferrite structure in each sample.

The magnetic properties of the systems were measured in a Vibrating Sample Magnetometer (VSM, Lakeshore) and in a commercial superconducting quantum interference device magnetometer (SQUID, MPMS Quantum Design). The saturation magnetization ($M_S$) was determined by magnetization measurements as a function of the applied field measured at 300 K in the VSM (see Supporting Information S3). The blocking temperature distribution ($f(T_B)$) of each sample was obtained from the magnetization curves as function of temperature measured in the zero-field-cooling (ZFC) and field-cooling (FC) protocols ($M_{ZFC}$ and $M_{FC}$ curves, respectively) in the SQUID magnetometer with an applied field of 4 kA/m. The blocking temperature distribution $f(T_B)$ is obtained by the expression $(1/T)d(M_{ZFC}(T)-M_{FC}(T))/dT$ (see Supporting Information S4), which resulted in a good agreement with blocking temperature distribution obtained from temperature decay of the remanence curve (TRM), as shown in a previous work [27]. Other approximations are used in the literature to obtain a blocking temperature [28-32] and the difference between the f($T_B$) obtained from these expressions depends of the dispersion of the size distribution; for comparison, one of them (f($T_B$) α d($M_{ZFC}$(T)-$M_{FC}$(T))/dT) is compared in Supporting Information S4. The samples for the magnetic measurements were conditioned by dispersing the NPs in epoxy resin in low concentration (about 0.1 wt.%), avoiding the agglomeration and physical rotation of the nanoparticles with the applied field.

In other to avoid the Thermal dependence of the specific heat ($c_p$(T)) of paraffin, CBO and CBO containing nanoparticles at concentrations of 0.2 wt.% (CBO+NPs) were determined by Differential Scanning Calorimetry (DSC) using a TA Instruments Q2000 calorimeter. A three-step procedure [33] has been performed at 10, 5 and 1 K/min, under 50 ml/min $N_2$ flow in hermetic aluminum crucibles, using sapphire as calibration substance. DSC results were analyzed in a similar way than used in [23] for CBO and in [16] for paraffin, reflecting the melting process. For CBO, two peaks were observed, one associated to the unresolved LMF and MMF melting and another peak related to the HMF melting. The last transition is the only one developed within the temperature range of hyperthermia experiments (292 K - 330 K). The enthalpy related to the last peak (Δ$H_m$) was determined as the area of the measured $c_p$(T) curve respect to the baseline curve ($c_p^{baseline}$(T)) as:

$$\Delta H_m \left[\frac{J}{g}\right] = \int_{T_i}^{T_F} c_p(T) dT - \int_{T_i}^{T_F} c_p^{baseline} dT, \qquad \text{eq. 1}$$

where $T_i$ and $T_F$ are the initial and final temperature of the HMF, respectively, defined by the linear extrapolations of the raising and the dropping in the initial and final steps of the HFM, respectively.

For paraffin, without nanoparticles, the melting process is related to a peak between 307.2 K ($T_i$) and 327.4 K ($T_F$).

The samples used in the MFH experiments were prepared by heating the CBO and paraffin until 317 K and 343 K, respectively, with a small amount of toluene solution containing the amount of NPs necessary to obtain the final concentration was dropped. Magnetization measurements as a function of the applied field measured at 300 K in the VSM of CBO and paraffin without nanoparticles are presented in Supplementary Information S5. For both, a diamagnetic component is observed indicating the absence of magnetic contamination in the detection limit of the experiment. This mixture was exposed to ultrasound during 30 minutes at 317 K for CBO and 343 K for paraffin, both in an open flask. MFH experiments were performed in two different commercial models (nB Nanoscale Biomagnetics, Spain), operating with magnetic field amplitude ($H_0$) of 16 kA/m: a DM100-model with working frequency (f) of 817 kHz and a F1-D5 RF-model with f = 570 kHz. These apparatus present different thermal insulation setups and magnetic field gradients produced by the solenoid, the DM100-model having better insulation and less field gradient. As calibration, the MFH experiments with CBO and paraffin without nanoparticles were performed from the room temperature, showing the absence of temperature increment in both cases (see Supplementary Information S6). The quantities $c_p(dT/dt)$ [W/g] and its integral in the corresponding time interval were numerically calculated with using an interpolation in the desired range for both curves after an advanced-averaging smoothing procedure for the measured T(t) curves, being labeled as $\Gamma(t)$:

$$\Gamma(t) \left[\frac{J}{g}\right] = \int_{t_0}^{t} c_p \left(\frac{dT}{dt}\right) dt. \qquad \text{eq. 2}$$

RESULTS AND DISCUSSION

Representative TEM images of synthesized NPs of the four samples are shown in Figure 1, with the respective histograms of diameters presented in the right panel and fitted with a lognormal distribution for all samples. The values of <d> and σ obtained from the fitting are given in Table I.

Concerning the magnetic properties of our samples, they present distinct mean blocking temperature $T_B$ (see Table 1) obtained from the $M_{ZFC}(T)$ and $M_{FC}(T)$ measurements presented in Figure 2. This is a consequence of the expected changes in the effective magnetic anisotropy $K_{eff}$ and volume of the samples. Figure 2 also shows the corresponding blocking temperature distribution $f(T_B)$ of all samples, according to the established in Supplementary Information S4, compared with the blocking temperature distribution $f(T_B)^*$ calculated from de diameter histogram obtained of TEM data (Figure 1) by the Néel's model with assuming a value of $\tau_0 = 10^{-10}$ s and a measuring time $\tau_m = 100$ s and with using the amplitude and the energy barrier as the fitting parameters. $K_{eff}$ was calculated through the definition of blocking temperature in Néel's model by matching d distribution from TEM results and the $f(T_B)$ distribution obtained from magnetic measurements. The obtained $K_{eff}$ values are presented in Table 1. A high irreversibility temperature is observed in all curves with exception of sample x = 0.19. A high irreversibility temperature is observed in all curves with exception of sample x = 0.19. These high irreversibility temperatures indicate the contribution of nanoparticles with large anisotropy but not so relevant in population.

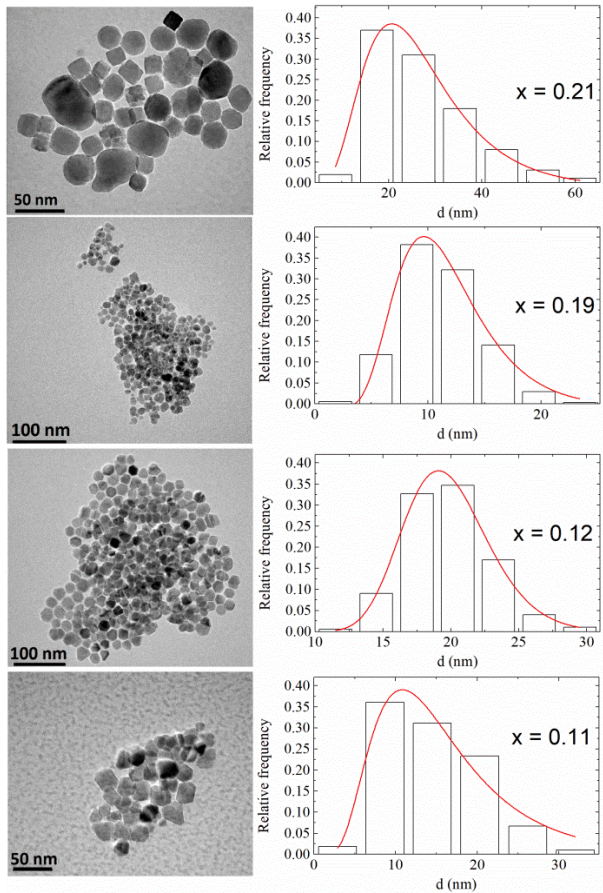

**Figure 1.** Representative TEM images of the as-prepared $Zn_xFe_{3-x}O_4$ nanoparticles. The right panel shows the respective diameter histograms fitted with a lognormal distribution whose parameters are given in Table I.

Table 1 also gives the saturation magnetization values ($M_S$) measured at 300 K for all samples. For the magnetization curves as function of applied field M(H), see Supporting Information S3.

Sample x = 0.19 presents a composition and morphology resulting in a product $K_{eff}d^3$ with relative narrow size dispersion close to a value that results in reasonable SPA values. For the other samples, the product $K_{eff}d^3$ would result in a reduced SPA value and, in this way, the broader size distributions of these samples result in a SPA that allows performing our MFH experiments in the temperature range of the phase transition in both media.

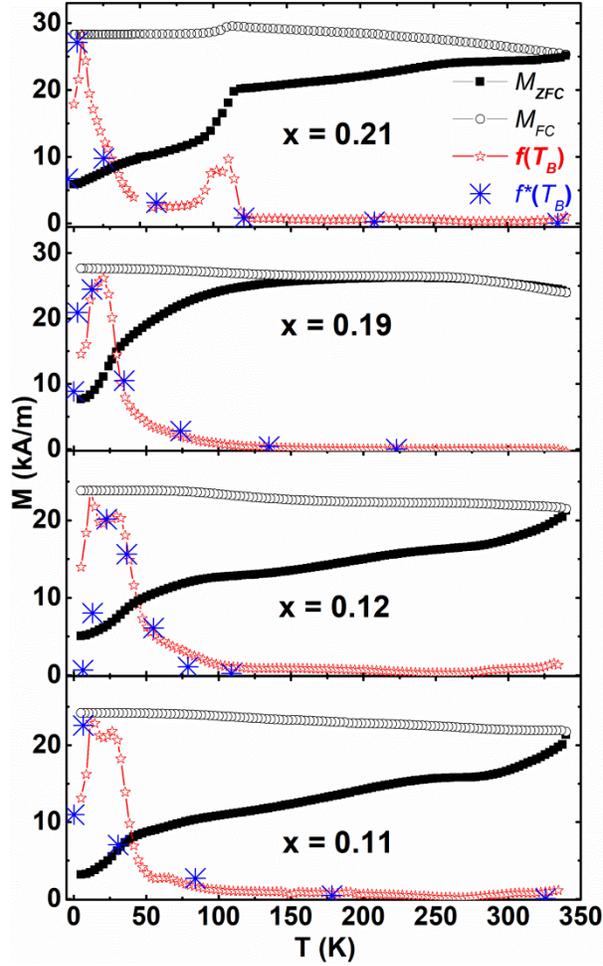

**Figure 2.** $M_{ZFC}(T)$ and $M_{FC}(T)$ curves of $Zn_xFe_{3-x}O_4$ NPs nanoparticles (solid and open squares, respectively) measured with an applied field of 4 kA/m. The respective blocking temperature distribution $f(T_B)$ (red stars) is also presented together with $f(T_B)^*$ (blue crosses) calculated from the size distributions given in Figure 1.

**Table 1.** Nanoparticle size distribution and magnetic parameters of $Zn_xFe_{3-x}O_4$ obtained from the TEM images and magnetization measurements.

| X | [1]<d> (nm) | [1]σ | [2]$M_S^{300K}$ (kA/m) | [3]<$T_B$> (K) | [4]$K_{eff}$ (J/m$^3$) |
|---|---|---|---|---|---|
| 0.21 | 24.7 | 0.41 | 74 | 55 | 3.0x10$^3$ |
| 0.19 | 11.0 | 0.18 | 75 | 32 | 2.5x10$^3$ |
| 0.12 | 19.7 | 0.36 | 69 | 72 | 5.2x10$^3$ |
| 0.11 | 14.0 | 0.50 | 63 | 64 | 7.5x10$^3$ |

[1]<d> is the mean particle size and σ is the size dispersion obtained from the fitting of the diameter histogram with a lognormal.

[2]$M_S^{300K}$ is the saturation magnetization at 300 K obtained from the hysteresis cycle.

[3]<$T_B$> is the mean blocking temperature obtained from the blocking temperature distribution f($T_B$) calculated from the $M_{ZFC}$ and $M_{FC}$ curves.

[4]$K_{eff}$ is the effective magnetic anisotropy obtained from the fit of f($T_B$) with the blocking temperature distribution calculated from de diameter histogram f($T_B$)*.

Figure 3(a) shows the results of the MFH experiment of all samples of NPs dispersed in CBO in a concentration of 0.2 wt.% performed with $H_0$ = 16 kA/m and f = 817 KHz from the initial temperature $T_0$ = 294.2 K in the DM100-model. As expected, the samples present distinct heating rate, which is consequence of different relaxation times resulting from different morphology, specifically <d> and σ; and composition, which reflects in different $M_S$ and anisotropy energy barrier proportional to <$T_B$>. Figure 3(b) gives the MFH results in the DM100-model ($H_0$ = 16 kA/m and f = 570 KHz) for the samples x = 0.21, 0.19, 0.12 and 0.11 dispersed in paraffin in a concentration of 0.4 wt.%, which present similar trends to the samples dispersed in CBO, excepting that the increment in the heating rate is observed in a higher temperature (~ 318 K).

Usually, Specific Power Absorption is obtained from the initial slope of these curves by the simplified relationship SPA = (m/$m_{NPs}$)$c_p$(dT/dt), where $c_p$ and m are the specific heat and the mass of the media, respectively, and $m_{NPs}$ is the mass of the nanoparticles. This expression assumes that all energy absorbed by the NPs from the *ac* applied field is used to increase the temperature and heat losses are negligible during the first few seconds of heating. After the initial seconds, however, the heating rate (dT/dt) is observed to depart from the linear behavior due to the non-adiabaticity of the system. However, in these systems of NPs dispersed in CBO and in paraffin, the heating curves reveal a complex behavior with regions with different heating rate, which increases around 306 K for all samples in CBO and around 316 K for paraffin, as indicated by the dotted lines in Figure 3(a) and Figure 3(b).

The increase in the heating rate during the MFH experiment observed in Figure 3(a) and Figure 3(b) could be due to changes in the thermodynamic properties of the media, or in a change of the nanoparticle relaxation mechanism, from Néel to Brown driven by the melting of the CBO (or paraffin) and the consequent reduction in the viscosity η of the CBO (in the case of paraffin a change from solid to liquid state). In order to evaluate the possibility of a change in the relaxation mechanism as a consequence of the CBO melting, we calculate the Brown ($\tau_B$) and Néel ($\tau_N$) relaxation times of all samples. The characteristic times were calculated as $\tau_B = 3\eta V_{hyd}/k_B T$ and $\tau_N = \tau_0 \exp(27<T_B>/T)$, by using $\tau_0 = 10^{-10}$ s, the volume obtained from TEM, the CBO viscosity η = 10 Pa.s and 10 mPa.s for T = 294 K and 312 K, lower and higher temperatures than the CBO melting point, respectively. We consider $V_{hyd} = (\pi/6)<d>^3$, without including an organic layer usually present in the nanoparticles prepared by this method and, therefore, the value of $\tau_B$ is underestimated. The relationship between $\tau_N$ and $\tau_B$ are plotted in the relaxation time diagrams build up according to Lima *et al.* [2] in Figure 3(c). From this figure, it is clear that $\tau_N < \tau_B$ (even considering an underestimated value of $\tau_B$) above and below the melting temperature. In this way, the magnetic relaxation mechanism is the dominant one for all samples and in all the studied temperature range. This result rule out the hypothesis that a change in the nanoparticles relaxation mechanism could be responsible for the increase in the heating rate in the MFH experiments during the CBO melting. Figure 3(d) give a similar diagram for the nanoparticles in paraffin ($\eta = 2.1$ mPa.s) for T = 353 K, resulting in $\tau_N < \tau_B$ for the three samples and reflecting the Néel relaxation mechanisms as the dominant one.

Figure 4(a) presents the $c_p(T)$ curves of CBO measured with different heating rates: 1 K/min and 10 K/min. As clearly observed, there are two peaks in both curves, which are better defined for the highest heating rate. The first peak is associated to the unresolved low-melting fraction and medium-melting fraction (LMF+MMF) transitions at lower temperature, and the second peak is associated to the high-melting fraction (HMF) transition of CBO [23]. The effect of total melting transition can be observed at plain eye, as shown in the pictures taken at different temperatures (see Supporting Information S7), where the melting is complete after the HMF transition. As mentioned before, the MFH experiments are performed in the temperature range where the HMF transition develops. The temperature range of the HMF peak ($\Delta T_M$) is well marked for both heating rates, and can be calculated from the extrapolated peak onset and offset temperatures in the $c^p(T)$ curves: 297.1 K to 310.8 K for 10 K/min and 299.8 K to 309.0 K for 1 K/min. For the heating rate of 10 K/min, where the peaks are better defined, a melting enthalpy value of 40.5 J/g is calculated by using eq. 1 and taking the area of both peaks, *i.e.*, for the whole melting process (LMF+MMF and HMF).

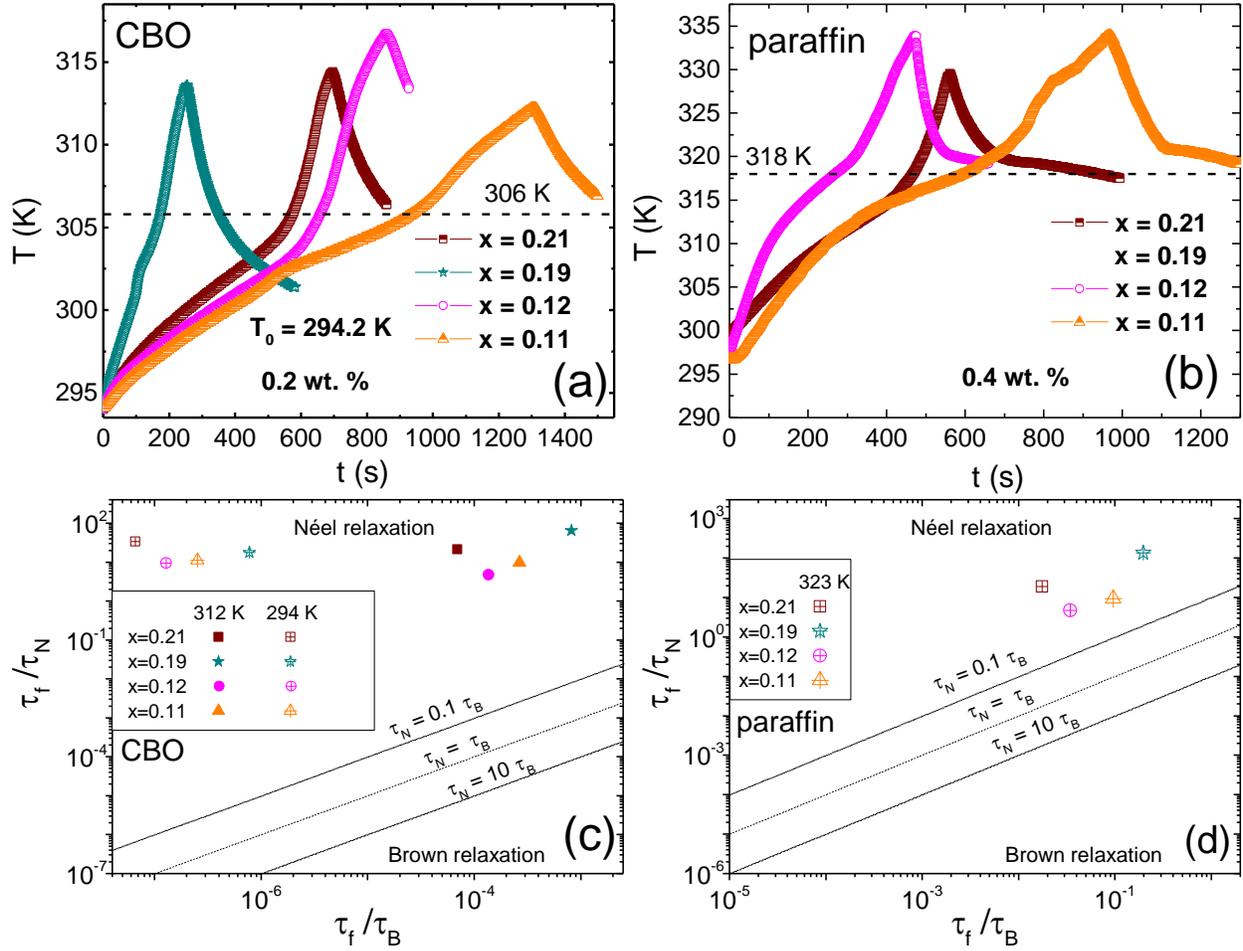

**Figure 3.** (a) and (b): Evolution of the temperature with the time during the MFH experiments performed in the DM100-model for 0.2 wt.% of samples x = 0.21, 0.19, 0.12 and 0.11 dispersed in CBO ($H_0$ = 16 kA/m and f = 817 kHz) and in paraffin ($H_0$ = 16 kA/m and f = 570 kHz), respectively. Dotted line indicates the temperature of 306 K and 318 K, where an increase in the slope is observed for CBO and paraffin, respectively. (c) and (d): Relaxation times diagram where the Néel ($\tau_N$) and Brown ($\tau_B$) relaxation times of the different nanoparticles dispersed in CBO and in paraffin, respectively, normalized by the applied field period ($\tau_f = 1/(2\pi f)$) are plotted; later data were calculated for two temperatures 294 K and 312 K, where the CBO presents different viscosities, and 323 K, above the melting of paraffin.

Figure 4(b) gives the $c_p(T)$ curves measured with heating rates of 1 K/min and 5 K/min for CBO containing x = 0.19 nanoparticles at the concentrations of 0.2 wt.% . As observed, these curves present differences among them and with respect to the curves of pure CBO. Notice that the presence of NPs results in smaller changes in the $\Delta T_M$ than those occasioned by different heating rates in pure CBO. For the heating rates of 1 and 5 K/min, $\Delta T_M$ of the HMF is between 297 K and 310 K. For the analysis of the MFH experiment, the $c_p(T)$ curves reported in Figure 4(a) and 4(b) were used for pure CBO and for CBO containing NPs, respectively. Finally, Figure 4(c) gives the $c_p(T)$ curve for the paraffin measured with heating rate of 10 K/min.

Independently from the effects of nanoparticles and heating rate, the important changes in the $c_p(T)$ curve of CBO should be taken into account for the analysis of the MFH results. In particular, the estimation of the SPA of the system should be revised when the specific heat presents a non-monotonous evolution, and latent heat is stored or released during the phase transition. In this way, the thermal energy of the system when the applied field is turned on and turned off can be written, respectively, as:

$$m_{CBO} c_P(T) dT = dQ_{SPA}(T) - dQ_{loss}(T) \quad (H_0 > 0) \quad \text{eq. 3}$$

$$m_{CBO} c_p(T) dT = -dQ_{loss}(T) \quad (H_0 = 0), \quad \text{eq. 4}$$

where $m_{CBO}$ is the mass of the medium (CBO), $Q_{SPA}$ is the heat released by the magnetic losses of the NPs for $H_0>0$ and $Q_{loss}(T)$ are the heat losses of the system due to the non-adiabatic condition of the MFH experimental apparatus [34]. The amount of $Q_{loss}$ depends on the MFH apparatus used. We remark again that, according to our measurements of cooling down curves (after the applied field is turned off) $Q_{loss}$ is lower for DM100-model than for F1-D5-model. In addition, $Q_{loss}(T)$ is expected to depend on the temperature difference between the sample and the environment. The melting enthalpy is considered in the thermal variation of $c_p(T)$ through eq. 1.

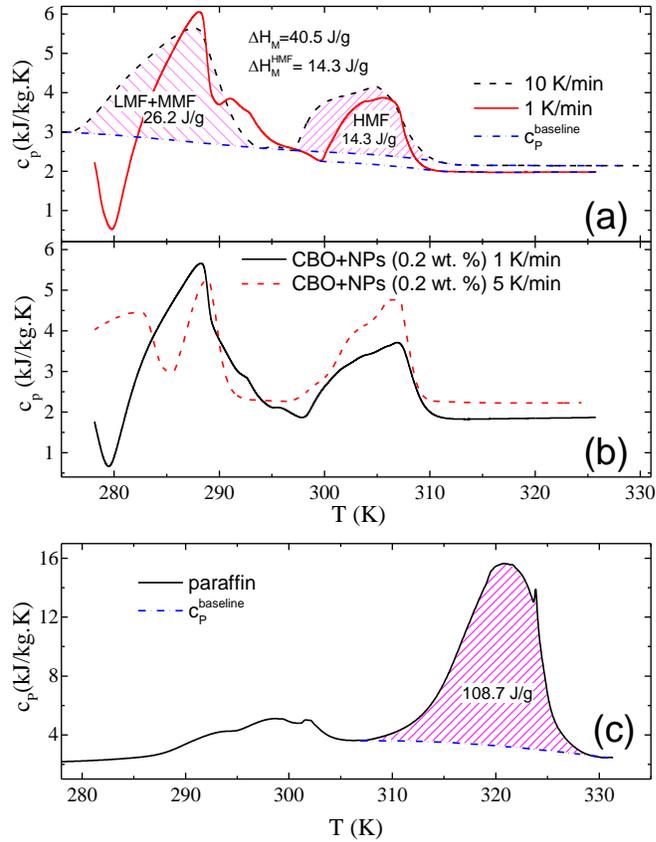

**Figure 4.** (a) Thermal dependence of specific heat ($c_p(T)$) of the CBO measured with heating rate of 1 and 10 K/min, as well as the $c_p^{baseline}(T)$ curves (dotted-blue line for both heating rates). Dashed zones indicates the difference between $c_p(T)$ and $c_p^{baseline}(T)$ associated to the melting of the distinct fractions. (b) $c_p(T)$ curves for CBO containing 0.2 wt.% of $x = 0.19$ nanoparticles with heating rates of 1 K/min and 5 K/min. (c) $c_p(T)$ curves for paraffin with heating rate of 10 K/min.

Taking into account the appropriated $c_p(T)$ curve together with eq. 3 and eq. 4, the hyperthermia curves could be divided in four regions: A, B and C, where the applied field is on ($H_0 > 0$), and D, where $H_0 = 0$. Figure 5 (a) shows these four regions in for sample $x = 0.21$ (concentration of 0.2 wt.%, $f = 817$ kHz and $H_0 = 16$ kA/m) using the DM100-model. Region A is in the temperature range below the HMF transition (T < 297 K), B coincides with the HFM melting (297 K < T < 309

K) and C is for temperatures above the HMF region (T > 309 K). As discussed previously, the raising in the MFH curve above 306 K could not be associated with a change in the dominant relaxation mechanism due to the strong reduction in the viscosity of CBO above the melting. In fact, a change in the CBO physical state is observed in the pictures for 307 K if compared to 305 K (see Supporting Information S7). Also, the reduction in the viscosity of CBO, some degrees below $T_F$ = 309 K, may induce some changes in the spatial arrangement of the nanoparticles, as discussed below.

Integration of eq. 3 and eq. 4 in the corresponding interval of temperature results:

$$\int_{T_0}^{T} c_p(T)dT = \int_{t_0}^{t} c_p(T)\left(\frac{dT}{dt}\right)dt = \frac{(Q_{SPA} - \Delta Q_{loss})}{m_{CBO}} \qquad (H_0 > 0) \qquad \text{eq. 5}$$

$$\int_{T_0}^{T} c_p(T)dT = \int_{t_0}^{t} c_p(T)\left(\frac{dT}{dt}\right)dt = \frac{-\Delta Q_{loss}}{m_{CBO}} \qquad (H_0 = 0). \qquad \text{eq. 6}$$

For eq. 5, the integral limits corresponding to the starting time $t_0$ and the final t where the field is turned off, with the corresponding temperatures $T_0$ and T. For eq. 6, these integral intervals represents the time and temperature ranges of region D, where $H_0 = 0$. Thus, eq. 5 gives information about the SPA in region A, B and C, while eq. 6 provides information of $\Delta Q_{loss}$.

Figure 5(b) gives the corresponding temporal dependence of $\Gamma(t)$ (eq. 2) calculated from the experimental data of MFH (figure 5(a)) and $c_p(T)$ with heating rate of 5 K/min (see figure 4(b)). This graph gives the accumulated thermal energy in CBO, normalized by its mass, due to the magnetic losses from the applied field and the heat loss due to non-adiabatic conditions. This curve is also divided in the four regions, taking the corresponding time when the corresponding temperature of figure 5a is reached. The raising above 306 K is also observed in these curves.

For region A, it is expected that $\Delta Q_{loss}$ has a relative low value in comparison to $Q_{SPA}$, since heat losses by the non-adiabatic condition are small at temperatures close to the ambient temperature ($T_e$ = 293.7 K). Thus, for region A, the SPA value can be approximated by the quantity ($m_{CBO}/m_{NPs}$)$c_p(T)(dT/dt)$. In fact, a relatively small decreasing of the SPA is observed with the increment of time, probably related with a small value $\Delta Q_{loss}$ that increases with increasing the temperature.

For region C this approach cannot be used since the temperature of the sample is higher than $T_e$ and $Q_{loss}$ should be taken into account. In this way, we propose a correction of the MFH temperature considering the $Q_{loss}$ obtained from the cooling curve, when the magnetic field is turned off in order to estimate the SPA in region C. The T(t) data in region D was fitted with the following expression that reflects its expected exponential decay with the time:

$$T(t) = T_e + \left[(T_{t0} - T_e)exp(-\beta(t - t_0))\right],  \qquad \text{eq. 7}$$

where $T_{t0}$ is the value of temperature at the time $t_0$ (moment when the field is turned off in this region) and β is the parameter that models the heating loss. $T_e$ is obtained from the low temperature range of the cooling curve of CBO containing 0.2 wt.% of NPs (see Supporting Information S8). The result of this fitting procedure is given in Figure 5(c). From this fitting, the corrected temperature, *i.e.*, the temperature expected for adiabatic condition, $T_{Adiab}$, is calculated from the measured temperature T(t) as:

$$T_{Adiab}(t) = T(t) + \beta \left[\int_{t_0}^{t} T(t)dT - T_e(t - t_0)\right]. \qquad \text{eq. 8}$$

Both $\Delta T_{Adiab}(t) = T_{Adiab}(t) - T_0(t)$ and $\Delta T(t) = T(t) - T_0(t)$ curves are presented in Figure 5(c) together with the fitting of the cooling region. As expected, $\Delta T_{Adiab}(t)$ presents an almost constant value in region D, as is expected for a system that does not exhibit losses. This result reinforces the fact that our temperature correction procedure is correct. In this way, the corrected value of SPA can be calculated for region C by ($m_{CBO}/m_{NPS}$)$c_p(T)(d\Delta T_{Adiab}(t)/dt)$.

For region B, fitting the SPA dependence with time is not simple because of the HMF melting within this temperature range. Nevertheless, it is possible to estimate a mean SPA value (<SPA>) by using eq. 5 as follow: i) the left side of eq. 5 in the time interval $T_i \leq T \leq T_F$ (see eq. 1) could be obtained directly from the data presented in figure 5b, as indicated, being 45.8 J/$g_{CBO}$; ii) the quantity $\Delta Q_{loss}$ can be estimated from the equivalent curve from the cooling of CBO+NPs until the temperature attain $T_e$ (see Supplementary Information S8), being 36.3 J/$g_{CBO}$. Therefore, taking into account these values, the mean value of <SPA> results 77 W/$g_{NPs}$. This <SPA> value is shown in Figure 5(d) together with the values of SPA obtained for regions A and C (here also included that obtained for $\Delta T(t)$ with the same expression, *i.e.*, uncorrected by $Q_{loss}$). Similar analysis was performed for samples x = 0.19 (Figures 5(e), 5(f), 5(g), and 5(h)), x = 0.12 (Figures 5(i), 5(j), 5(k) and 5(l)) and x = 0.11 (Figures 5(m), 5(n), 5(o) and 5(p)). Sample x = 0.19 presents a higher heating rate in MFH experiments when compared with sample x = 0.21, resulting in less evident changes: the decreasing of SPA in region A with increasing the temperature and a less pronounced decrease in the heating rate between $T_i$ = 297 K. In opposition, samples x = 0.12 and x = 0.11 have a smaller heating rate in comparison to sample x = 0.21, making more pronounced the changes mentioned above.

The <SPA> values estimated for region B has an intermediate value between those obtained for regions A and C in all samples. Probably, the SPA is not constant along this region, with a probable increment above 306 K. The decreasing in the heating rate of MFH curves between 297 K and 306 K could be probably associated to two contributions. Firstly, the increase of $\Delta Q_{loss}$ with respect to the $Q_{SPA}$ with increasing in the temperature, which could be observed in the decreasing of the SPA with increasing the temperature in region A being more evident for samples with lower SPA. Secondly, the latent heat of the CBO melting reduces the effects of $Q_{SPA}$ in the temperature increment. It is more evident for the samples with smaller $Q_{SPA}$.

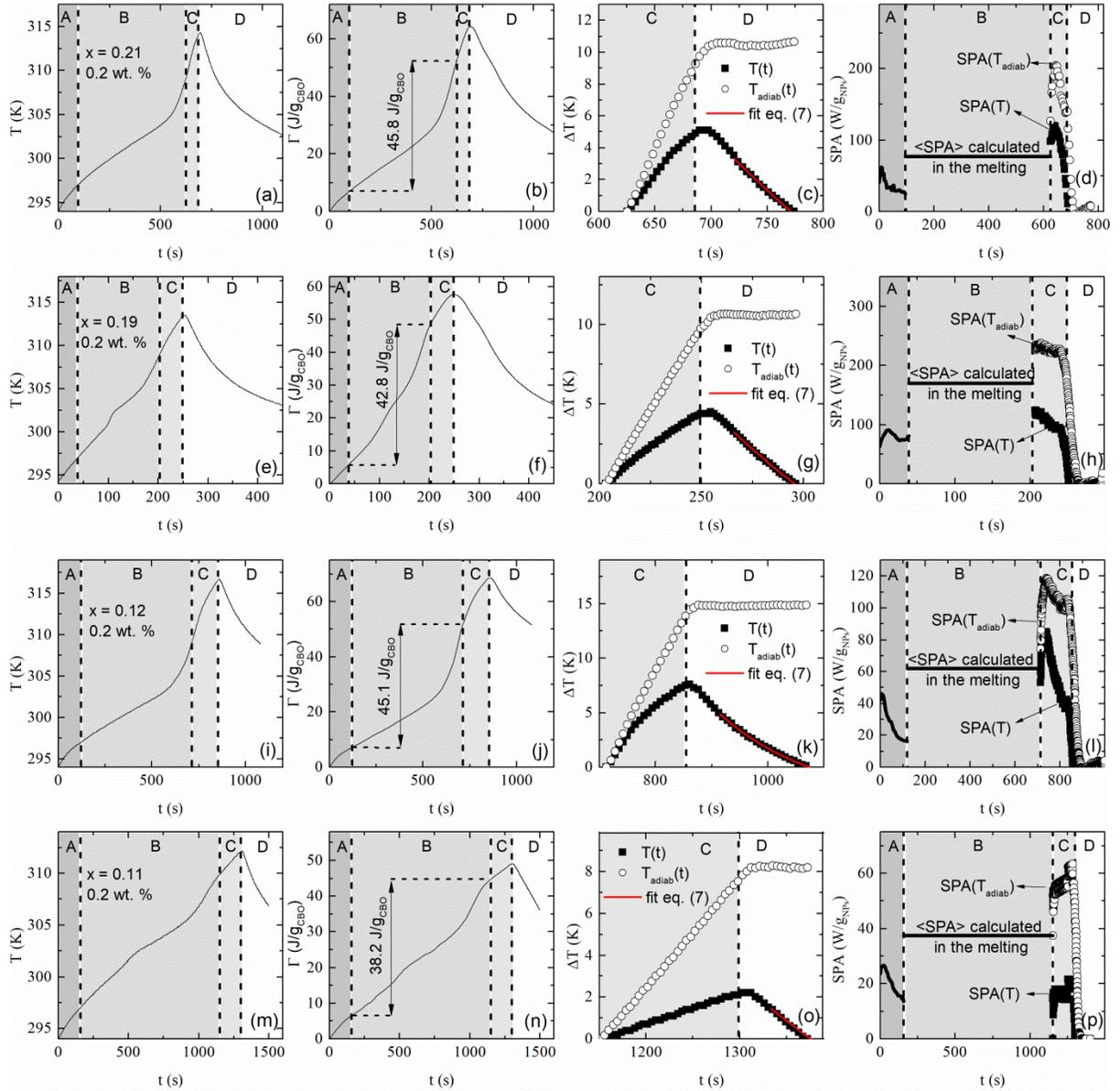

**Figure 5.** (a) T(t) results of MFH experiments for CBO containing 0.2 wt.% of x = 0.21 NPs with f = 817 kHz and $H_0$ = 16 kA/m in DM100-model. Regions A, B and C correspond to the temperature range defined below, within and above the HMF peak in $c_P(T)$ curve associated to the melting of CBO+NPs. (b) Γ(t) curves obtained from the MFH results and the line gives the heat variation within the interval $T_i$ and $T_F$. (c) ΔT(t)=T(t)-$T_0$ curve measured for region C fitted with eq. 7 and the temperature corrected for adiabatic condition (ΔT$_{Adiab}$(t)=T$_{Adiab}$(t)-$T_0$) calculated with eq. 8. (d) Estimated SPA values for regions A and C, as well the mean SPA (<SPA>) estimated for region B.

Similar curves for samples x = 0.19, x = 0.12 and x = 0.11 in the same concentration are presented in (e), (f), (g), (h), (i), (j), (k), (l), (m), (n), (o) and (p), respectively.

The SPA in region C increases between 2.5 and 3 times for all samples with respect to the values in region A, corresponding to the clearly noticed increment in the heating rate observed in the T(t) curves measured in MFH experiments. This increment in the SPA for region C is reflected in the increased slope in the MFH results above 306 K, some degrees below the ending of the melting and corresponding to a physical change in the CBO, with a strong reduction in the viscosity. Despite the reduction of the viscosity with the melting, $\tau_N$ should be several orders of magnitude smaller than $\tau_B$ (see Figure 3(b)), which is reinforced by the agglomeration process discussed below. However, the phase transition of the CBO with the consequent reduction in the viscosity allows other changes in the sample with consequences in the magnetic response of the system, as discussed below. The SPA in region C remains almost constant until the applied field is turned-off for samples x = 0.19 and x = 0.11. For the samples x = 0.21 and x = 0.12, the increment in the SPA is slightly slower and a decreasing in the SPA value is observed before the applied field is turned off. Interestingly, the SPA in region C is almost constant for the smaller nanoparticles, and it decrease with time for the two larger ones, which could be also related to an agglomeration or precipitation phenomena with the reduction in the viscosity of CBO above the melting.

A reproducible behavior was observed in the T(t) curves after three successive heating cycles within $T_0 < T \leq 305.5$ K and subsequent cooling (Figure 6(a), runs 1, 2 and 3) in the DM100-model (f = 817 kHz and $H_0$ = 16 kA/m) for CBO containing 0.15 wt.% of sample x = 0.19. On the other hand, heating the samples above the melting region resulted in a clear non-reproducible data (runs 4 and 5 in Figure 6(b)). In can be noted that the T(t) curve for run 4 is similar to run 3 up to T ~ 301

K, whereas in run 5 deviations from run 3 are observed even below that temperature. We associate this change in the heating of the system to the agglomeration process of the nanoparticle in the melted CBO, less viscous, probably induced by the magnetic field gradient inside the coil and, to a lesser extent, to gravity. Different spatial configuration and NPs aggregation degree after successive runs could change the SPA, as observed in the analysis presented in Figure 5, and it could also lead to the observation of hysteresis behavior. In fact, the formation of large asymmetric agglomerates (not-planar or bulk) is clearly observed at naked eye after the successive runs (see Figure 6(b)), which is not observed after successive runs when the heating temperature is kept below 305.5 K (see inset of Figure 6(a)). The irreversible effect in MFH curves becomes reversible if the particles are dispersed again, for example with ultrasound exposition, as shown in Figure 6(c) for three successive runs of the same sample measured in the D5-F1-Model (f = 570 kHz and $H_0$ = 16 kA/m) with intermediate exposition to the ultrasound (about 20 min).

Figure 7 (a-p) present a similar analysis to that of Figure 5 for samples x = 0.21, 0.19, 0.12 and 0.11 dispersed in paraffin in a concentration of 0.4 wt.% with $H_0$ = 16 kA/m and f = 570 kHz.. The graphics present general tendencies similar to those ones for the sample dispersed in CBO, excepting the higher temperature for the observation of the increment in the heating rate and the effects of the phase transition. In a similar way that the sample dispersed in CBO, Figure 8(a) shows the reversibility of the hyperthermia results for sample x = 0.11 dispersed in paraffin measured in the D5-F1-Model (f = 570 kHz and $H_0$ = 16 kA/m) when the heating rate is below 316 K (runs 1-3). Images taken (inset of Figure 8(a)) before and after each run evidence that no changes are observed in the temperature. However, when the temperature overcomes the melting point (runs 4 and 5 in Figure 8(b)), the nanoparticles agglomerate (see the inset of Figure 8(b)) and irreversibility is obtained, similarly to the observed for the CBO. Interesting, the increment in the heating rate is not observed when the nanoparticles are dispersed in toluene (see Supplementary Information S10), without a phase transition in the temperature range of the MFH experiment.

Considering that the change in the MFH curves above the melting temperature can be addressed to this agglomeration process, some speculation can be made with respect to the agglomeration process. There is currently some debate about the effects of particle agglomeration on the heating mechanisms. While some works reported a reduction of the SPA due to the formation of bulk agglomerates [6,9], there is also experimental evidence that the formation of elongated agglomerates (*e.g.* needle-like, cylinder or chain-like structures) increases the SPA compared to the well disperse NPs [12-14]. Recently, some theoretical models and experimental work showed that the actual effects of clustering could be either way, depending of the value of the effective magnetic anisotropy and anisotropy energy barrier of the NPs [35,36]. Basic considerations from magnetostatics indicate that in the presence of a non-uniform magnetic field the NPs tend to form elongated structures aligned with the magnetic field gradient. Probably, in our work, the reduction in the viscosity of the CBO above the melting allows the alignment of the easy axis of the NPs with the applied field at the same time that not-planar structures, probably elongated ones, are induced by the combination of the magnetic field gradient inside the coil and, to a lesser extent, to gravity. Any real applicator solenoid (even in a perfectly constructed one) will contain a non-uniform magnetic field along the coil axis, due to finite size effects. The formation of elongated structures after the first heating cycle, and the consequent increase in SPA in our case, is consistent with this explanation. According to this, the resulting heating curve will therefore depend on the magnetic field gradient (a characteristic of each MFH apparatus) and the NPs concentration due to the inter-particle interaction strength.

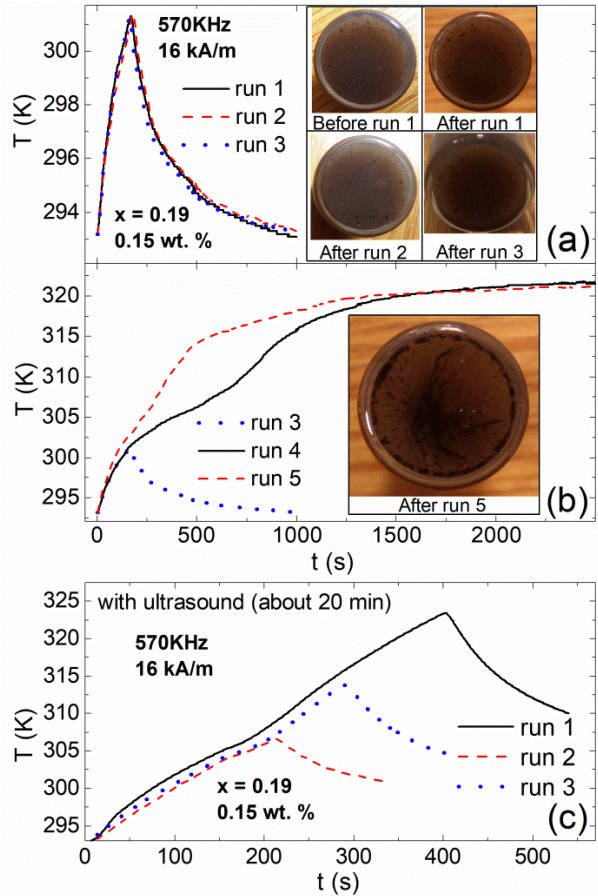

**Figure 6.** (a) MFH results for three subsequently runs performed in the DM100-model with f = 570 kHz and $H_0$ = 16 kA/m for CBO containing 0.15 wt.% of x = 0.19 NPs from $T_0$ to a final temperature below 306 K, followed by the cooling to $T_0$. (b) Two subsequently runs (4 and 5) measured in the same experimental conditions and performed from $T_0$ to a final temperature above the melting temperature (T > $T_F$). Insets present the picture of the samples after the indicated run. (c) Three successive runs of the same sample measured in the D5-F1-Model (f = 570 kHz and $H_0$ = 16 kA/m) with intermediate exposition to the ultrasound (about 20 min).

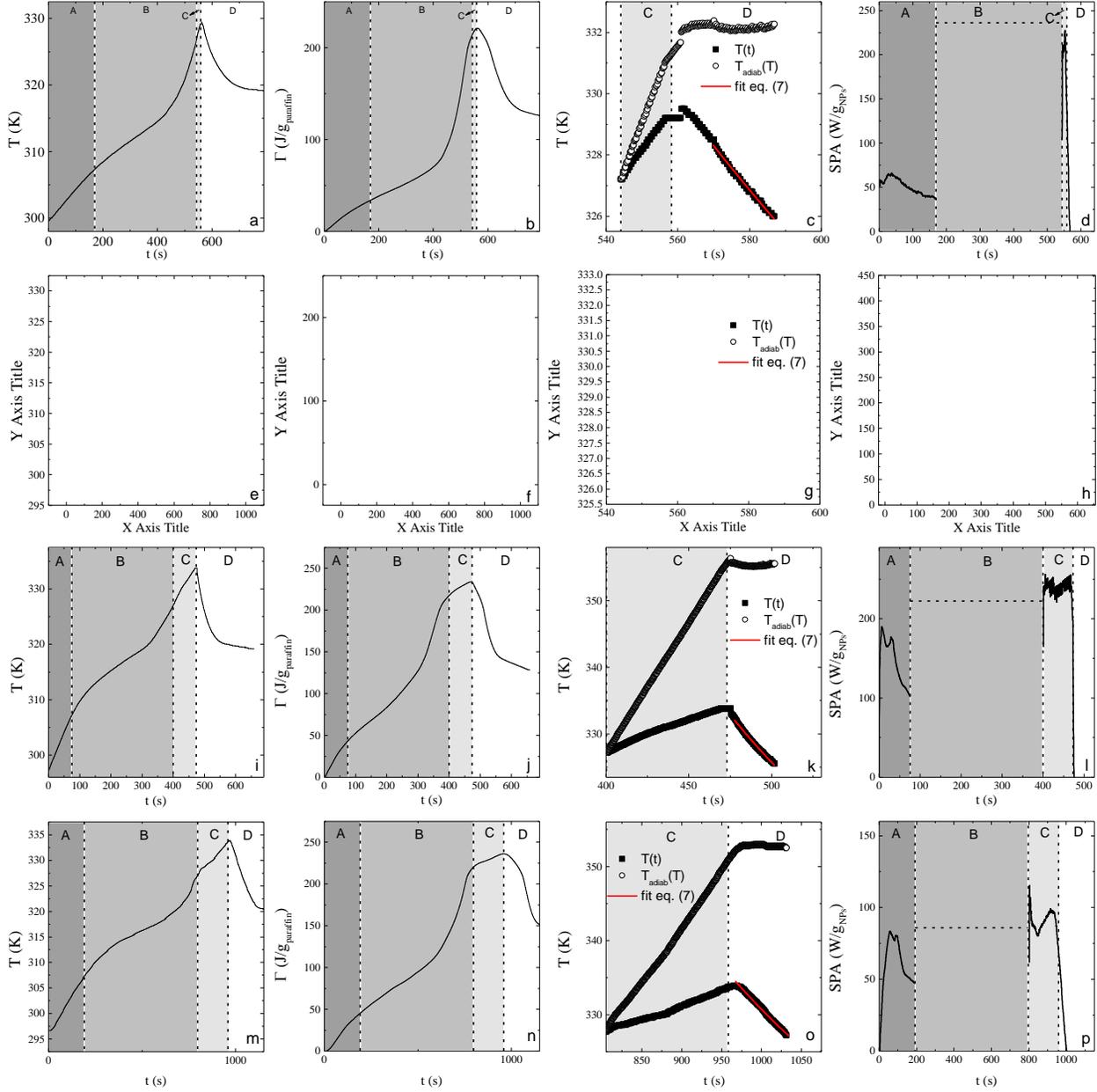

**Figure 7.** (a) T(t) results of MFH experiments for paraffin containing 0.4 wt.% of x = 0.21 NPs with f = 570 kHz and $H_0$ = 16 kA/m in DM100-model. Regions A, B and C correspond to the temperature range defined below, within and above the melting peak in $c_p(T)$ curve. (b) Γ(t) curves obtained from the MFH results and the line gives the heat variation within the interval $T_i$ and $T_F$. (c) ΔT(t)=T(t)-$T_0$ curve measured for region C fitted with eq. 7 and the temperature corrected for adiabatic condition calculated with eq. 8. (d) Estimated SPA values for regions A and C, as well the mean SPA (<SPA>) estimated for region B. Similar curves for samples x = 0.19, x = 0.12 and x = 0.11 in the same concentration are presented in (e), (f), (g), (h), (i), (j), (k), (l), (m), (n), (o) and (p), respectively.

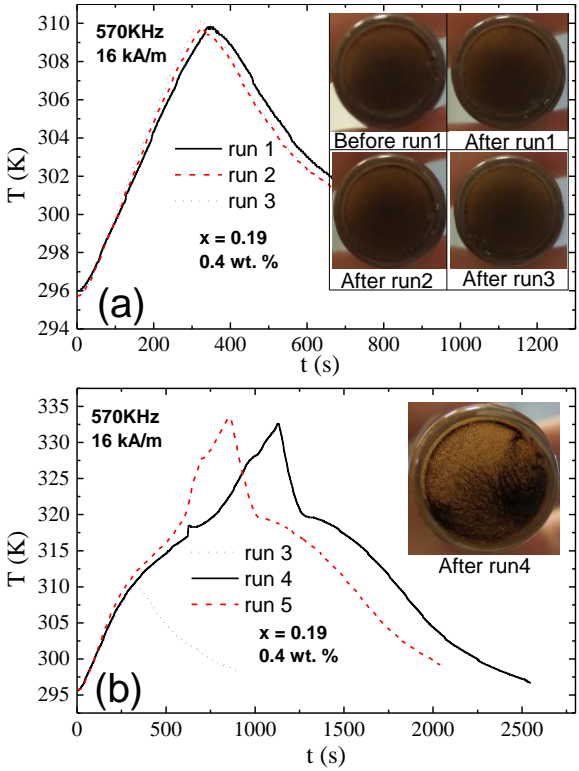

**Figure 8.** (a) MFH results for three subsequently runs performed in the D5-F1-Model with f = 570 kHz and $H_0$ = 16 kA/m for paraffin containing 0.4 wt.% of x = 0.11 NPs from $T_0$ to a final temperature below 316 K, followed by the cooling to $T_0$. (b) Two subsequently runs (4 and 5) measured in the same experimental conditions and performed from $T_0$ to a final temperature above the melting temperature (T > $T_F$). Insets present the picture of the samples after the indicated run.

The results presented in this work call the attention on the complex heat transfer process in MFH experiments from the NPs to the medium, where the thermodynamic characteristic of the media should be taken into account to evaluate the efficiency of nanoparticles systems in MFH treatments when they are applied in different environment. In particular the CBO and paraffin are used to emulate high viscosity environment and to evaluate the efficiency of the MFH when the Néel relaxation is the dominant mechanism, evidencing how the intrinsic transition of the media could affect the heating in a MFH experiment. Several aspects as the presence of latent heat in a transition, the strong variation of the specific heat with temperature and the formation of nanoparticles agglomerates when the viscosity diminishes after the melting are some of the aspects that should be evaluated in different media with complex structure or complex thermal response to analyze MFH results and the therapy effectiveness.

CONCLUSIONS

The Specific Power Absorption of $Zn_xFe_{3-x}O_4$ magnetic nanoparticles dispersed in CBO and paraffin present a complex behavior with temperature related to the melting of these media. Samples present distinct morphology and magnetic properties, specifically distinct anisotropy energy barrier. For all samples, the magnetic relaxation mechanism is dominated by Néel relaxation, independently of the CBO or paraffin phase, below and above the melting. The complex SPA behavior with temperature for this system is addressed to the non-linear thermal response of viscosity and thermodynamic properties of the medium, including a phase transition in the temperature range of MFH experiment. The variation of the $c_p$ of the medium with the temperature, as well as the latent heat of the melting is important to analyze the SPA evolution of all samples. The heating rate strongly depends on the variation of these thermodynamic properties.

Despite the change in the viscosity of the media with the melting is not high enough to change the dominant relaxation mechanism; it allows the nanoparticle agglomeration induced by the applied field gradient of the MFH coil, which leads to changes in the SPA when the system is heated above the melting. This effect in the SPA depends on the nanoparticle concentration and it also results in a hysteresis in MFH experiments for subsequent runs when the melting temperature is attained.

Finally this work highlights the importance of measuring the thermodynamic properties of the media where the nanoparticles would be applied, in order to evaluate the efficiency of the MFH treatment, especially in complex media where different transitions could take place.

ACKNOWLEDGMENT

The authors acknowledge financial support of Argentinian governmental agency ANPCyT (Project No.PICT-2016-0288 and PICT-2015-0883) and UNCuyo (Project No.06/C527 and 06/C528). The authors gratefully acknowledge the EU-commission financial support under the: H2020-MSCA-RISE-2016 SPICOLOST PROJECT No 734187. GFG and TET also acknowledge the financial support from the Spanish Ministerio de Ciencia, Innovación y Universidades (project MAT2016-78201-P) and the Aragon Regional Government (DGA, Project No. E26).

**Supporting Information**.

S1: Compositional analysis of the four samples with Proton Induced X-ray Emission (PIXE).

S2: X-ray diffraction (XRD) measurements of sample X=0.21 as representative of the samples.

S3: Saturation Magnetization ($M_S$) determined by from the magnetic measurements (M(H) curves).

S4: Obtaining the blocking-temperature distribution $f(T_B)$ from the magnetization measurements as function of temperature ($M_{ZFC}(T)$ and $M_{FC}(T)$).

S5: Magnetization as function of applied field (*M(H)* curves) measured in VSM for CBO and paraffin without nanoparticles at 300 K.

S6: Magnetic Fluid Hyperthermia measurements of CBO and Paraffin without Nanoparticles performed in the F1-D5 RF-model.

S7: Changes in the physical aspects of Clarified Butter Oil (CBO) with the temperature, which is related with the changes in the viscosity and specific heat.

S8: Cooling curve of CBO+NPs heated in a hot plate up to a temperature above the melting taken in the DM100-applicator and its analysis with respect to the quantity ($\Gamma(t)$) in order to estimate $\Delta Q_{loss}$.

S9: Cooling curve of paraffin with NPs heated in a hot plate up to a temperature above the melting taken in the DM100-applicator and its analysis with respect to the quantity ($\Gamma(t)$) in order to estimate $\Delta Q_{loss}$.

S10: Magnetic Fluid Hyperthermia (MFH) of the four samples dispersed in toluene (0.4 wt.%) performed in the DM100-applicator with amplitude of 16 kA/m and frequency of 570 kHz.

REFERENCES

[1] R. E. Rosensweig. Heating magnetic fluid with alternating magnetic field. J. Magn. Magn. Mater. **252**, 370 (2002).

[2] E. Lima Jr., E. De Biasi, R. D. Zysler, M. Vasquez Mansilla, M. L Mojica-Pisciotti, T. E. Torres, M. Pilar Calatayud, C. Marquina, M. R. Ibarra and G. F. Goya. Relaxation time diagram for identifying heat generation mechanisms in magnetic fluid hyperthermia. J. Nanopart. Res. **16**, 2791 (2014).

[3] F. Fabris, E. Lima, Jr., E. De Biasi, H. E. Troiani, M. Vasquez Mansilla, T. E. Torres, R. Fernández-Pacheco, M. R. Ibarra, G. F. Goya, R. Zysler et al. Controlling the dominant magnetic relaxation mechanisms for magnetic hyperthermia in bimagnetic core–shell nanoparticles. Nanoscale **11**, 3164 (2019).

[4] J. Carrey, B. Mehdaoui and M. Respaud. Simple models for dynamic hysteresis loop calculations of magnetic single-domain nanoparticles: Application to magnetic hyperthermia optimization. J. Appl. Phys. **109**, 083921 (2011).

[5] T. E. Torres, E. Lima Jr., M. P. Calatayud, B. Sanz, A. Ibarra, R. Fernández-Pacheco, A. Mayoral, C. Marquina, M. R. Ibarra and G. F. Goya. The relevance of brownian relaxation as power absorption mechanism in magnetic hyperthermia. Sci. Rep. **9**, 3992 (2019).

[6] E. Lima Jr., E. De Biasi, M. Vasquez Mansilla, M. E. Saleta, M. Granada, H. E. Troiani, F. B. Effenberger, L. M. Rossi, H. R. Rechenberg and R. D. Zysler. Heat generation in agglomerated ferrite nanoparticles in an alternating magnetic field. J. Phys. D: Appl. Phys. **46**, 045002 (2013).

[7] B. Sanz, M. P. Calatayud, E. De Biasi, E. Lima Jr., M. Vasquez Mansilla, R. D. Zysler, M. R. Ibarra and G. F. Goya. In silico before in vivo: how to predict the heating efficiency of magnetic nanoparticles within the intracellular space. Sci. Rep. **6**, 38733 (2016).

[8] M. E. Sadat, R. Patel, J. Sookoor, S. L. Bud'ko, R. C. Ewing, J. Zhang, H. Xu, Y. Wang, G. M. Pauletti, D. B. Mast et al. Effect of spatial confinement on magnetic hyperthermia via dipolar interactions in Fe3O4 nanoparticles for biomedical applications. Mater. Sci. Eng. C **42**, 52 (2014).

[9] L. C. Branquinho, M. S. Carrião, A. S. Costa, N. Zufelato, M. H. Sousa, R. Miotto, R. Ivkov and A. F. Bakuzis. Effect of magnetic dipolar interactions on nanoparticle heating efficiency: implications for cancer hyperthermia. Sci. Rep. **3**, 2887 (2013).

[10] D. F. Coral, P. Mendoza Zélis, M. Marciello, M. del Puerto Morales, A. Craievich, F. H. Sánchez and M. B. Fernández van Raap. Effect of nanoclustering and dipolar interactions in heat generation for magnetic hyperthermia. Langmuir **32**, 1201 (2016).

[11] R. Hiergeist, W. Andrä, N. Buske, R. Hergt, I. Hilger, U. Richter and W. Kaiser. Application of magnetite ferrofluids for hyperthermia. J. Mag. Mag. Mater. **201**, 420 (1999).

[12] B. Mehdaoui, R. P. Tan, A. Meffre, J. Carrey, S. Lachaize, B. Chaudret and M. Respaud. Increase of magnetic hyperthermia efficiency due to dipolar interactions in low-anisotropy magnetic nanoparticles: Theoretical and experimental results. Phys. Rev. B **87**, 174419 (2013).

[13] C. Martinez-Boubeta, K. Simeonidis, A. Makridis, M. Angelakeris, O. Iglesias, P. Guardia, A. Cabot, L. Yedra, S. Estradé, F. Peiró et al. Learning from nature to improve the heat generation of iron-oxide nanoparticles for magnetic hyperthermia applications. Sci. Rep. **3**, 1652 (2013).

[14] D. Serantes, K. Simeonidis, M. Angelakeris, O. Chubykalo-Fesenko, M. Marciello, M. del Puerto Morales, D. Baldomir and C. Martinez-Boubeta. Multiplying magnetic hyperthermia response by nanoparticle assembling. J. Phys. Chem. C **118**, 5921 (2014).

[15] A. R. Vakhshouri. Paraffin as phase change material, in: Paraffin, an overview, Ed. F. S. Soliman, IntechOpen-London, Chap. 5 (2020).

[16] P. S. Jesumathy. Latent heat thermal energy storage system. in: Paraffin, an overview, Ed. F. S. Soliman. IntechOpen-London, Chap. 4 (2020).

[17] B. M. Mehta. Butter, butter oil and ghee. in: Gourmet and health-promoting specialty oils, Ed. R. A. Moreau, A. Kamal-Eldin, AOCS Press-Urbana, Illinois, Chap. 21, pp. 527 (2009).

[18] S. Rønholt, P.Buldo, K. Mortensen. U. Andersen, J. C. Knudsen and L. Wiking. The effect of butter grains on physical properties of butter-like emulsions. J. Dairy Sci. **97**, 1929 (2014).

[19] S. P. Changade, R. V. Tambat and R. R. Kanoje. Physical properties of ghee prepared from high acidic milk-II. J. Dairying, Foods & H.S. **25**, 101 (2006).

[20] A. Devl and B. Khatkar. Thermo-physical properties of fats and oils. IJETR **7**, 45 (2017).

[21] N. Duhan, J. K. Sahu and S. N. Naik. Temperature dependent steady and dynamic oscillatory shear rheological characteristics of indian cow milk (Desi) ghee. J. Food Sci. Technol. **55**, 4059 (2018).

[22] E. L. Watson. Thermal properties of butter. Can. Agr. Eng. **17**, 68 (1975).

[23] J. Tomaszewska-Gras. Melting and crystallization DSC profiles of milk fat depending on selected factors. J. Therm. Anal. Calorim. **113**, 199 (2013).

[24] J. Lohr, A. A. de Almeida, M. S. Moreno, H. E. Troiani, G. F. Goya, T. E. Torres, R. Fernández-Pacheco, E. L. Winkler, M. Vasquez Mansilla, R. Cohen et al. Effects of Zn substitution in the magnetic and morphological properties of Fe-oxide-based core–shell nanoparticles produced in a single chemical synthesis. J. Phys. Chem. C **123**, 1444 (2019).

[25] S. A. E. Johansson, J. L. Campbell and K. G. Malmqvist. Particle-induced x-ray emission spectrometry (PIXE), J. D. Winefordner (Ed.), John Wiley & Sons (1995).

[26] J. L. Campbell, N. I. Boyd, N. Grassi, P. Bonnick and J. A. Maxwell. The Guelph PIXE software package IV. Nucl. Instrum. Methods Phys. Res., B **268**, 3356 (2010).

[27] G. C. Lavorato, E. Lima Jr., D. Tobia, D. Fiorani, H. E. Troiani, R. D. Zysler and E. L. Winkler. Size effects in bimagnetic CoO/CoFe2O4 core/shell nanoparticles. Nanotechnology **25**, 355704 (2014).

[28] H. Mamiya, M. Ohnuma, I. Nakatani and T. Furubayashim. Extraction of blocking temperature distribution from zero-field-cooled and field-cooled magnetization curves. IEEE T. Magn. **41**, 3394 (2005).

[29] D. A. Balaev, S. V. Semenov, A. A. Dubrovskiy, S. S. Yakushkin, V. L. Kirillov, O. N. Martyanov. Superparamegntic blocking in an ensemble of magnetite nanoparticles upon interparticle interactions. J. Magn. Magn. Mater. **440**, 199 (2017).

[30] N. A. Usov. Numerical simulation of field-cooled and zero field-cooled processes for assembly of superparamagnetic nanoparticles with uniaxial anisotropy. J. Appl. Phys. **109**, 023913 (2011).

[31] K. L. Livesey, S. Ruta, N. R. Anderson, D. Baldomir, R. W. Chantrell and D. Serantes. Beyond the blocking model to fit nanoparticle ZFC/FC magnetization curves. Sci. Rep. **8**, 11166 (2018).

[32] G. Muscas, G. Concas, S. Laureti, A. M. Testa, R. Mathieu, J. A. De Toro, C. Cannas, A. Musinu, M. A. Novak, C. Sangregorio et al. The interplay between single particle anisotropy and interparticle interactions in ensembles of magnetic nanoparticles. Phys. Chem. Chem. Phys. **20**, 28634 (2018).

[33] G. W. H. Höhne, W. F. Hemminger and H. -J. Flammersheim. Differential scanning calorimetry, Springer, Berlin (2003).

[34] E. Natividad, M. Castro and A. Mediano. Adiabatic vs. non-adiabatic determination of specific absorption rate of ferrofluids. J. Magn. Magn. Mater. **321**, 1497 (2009).

[35] A. F. Abu-Bakr and A. Yu. Zubarev. On the theory of magnetic hyperthermia: clusterization of nanoparticles. Phil. Trans. R. Soc. A. **378**, 20190251 (2020).

[36] E. Myrovali, N. Maniotis, A. Makridis, A. Terzopoulou, V. Ntomprougkidis, K. Simeonidis, D. Sakellari, O. Kalogirou, T. Samaras, R. Salikhov et al. Arrangement at the nanoscale: effect on magnetic particle hyperthermia. Sci. Rep. **6**, 37934 (2016).